\documentclass[aps,pra,preprintnumbers,amsmath,amssymb,twocolumn,
epsf,superscriptaddress,groupedaddress,nofootinbib,tightenlines]{revtex4}

\usepackage{graphicx}
\usepackage{amsmath}
\usepackage{bm}
\usepackage{color}
\usepackage{hyperref}

\def\beqn{\begin{eqnarray}}
\def\eeqn{\end{eqnarray}}
\def\barr{\begin{array}}
\def\earr{\end{array}}
\def\btab{\begin{tabular}}
\def\etab{\end{tabular}}
\def\bite{\begin{itemize}}
\def\eite{\end{itemize}}
\def\bcen{\begin{center}}
\def\ecen{\end{center}}

\def\eq{\begin{equation}}
\def\ee{\end{equation}}
\def\nn{\nonumber}

\newcommand{\ba}{\begin{eqnarray}}
\newcommand{\ea}{\end{eqnarray}}

\begin{document}
\title{
$\mu-H$ Lamb shift:  
dispersing the nucleon-excitation uncertainty \\
with a finite energy sum rule.
}
\author{Mikhail Gorchtein}
\affiliation{{Contact author;} 
Institut f\"ur Kernphysik, Universit\"at Mainz, 55128 Mainz, Germany}
\email{gorshtey@kph.uni-mainz.de}
\author{Felipe J. Llanes-Estrada}
\affiliation{Dept. Fisica Teorica I, Universidad Complutense de Madrid, Madrid 28040, Espa\~na}
\author{Adam P. Szczepaniak}
\affiliation{
 Department of Physics  and Center for Exploration of Energy and Matter, Indiana University, Bloomington, IN 47403 USA} 
\begin{abstract}
We assess the two-photon exchange contribution to the Lamb shift in
muonic hydrogen with forward dispersion relations. 
The subtraction constant $\bar T(0,Q^2)$ that is necessary for a
dispersive evaluation of the forward doubly-virtual Compton amplitude,
through a finite energy sum rule, is related to the fixed $J=0$ pole
generalized to the case of virtual photons.  We evaluated this sum
rule using excellent virtual photoabsorption data that are available.
We find that the ``proton polarizability correction'' 
to the Lamb shift in muonic hydrogen is $-(40\pm5)\mu$eV. 
We conclude that nucleon structure-dependent uncertainty by itself is 
unlikely to resolve the large (300$\mu$eV) discrepancy between 
direct measurement of the Lamb shift in $\mu H$ and expectations 
based on conventional Hydrogen measurements.  
\end{abstract}
\maketitle
\section{Introduction} 

An ongoing controversy surrounding the proton size originates from the large discrepancy between the recent measurement of the Lamb shift in muonic hydrogen and earlier measurement based on conventional hydrogen as well as electron scattering (see for example the review~\cite{Jentschura:2010ej}). The advantage of using the muonic hydrogen over the conventional is that due to a larger reduced mass  the Lamb shift in the former is by an order of magnitude more sensitive to the proton radius. 
 The Lamb shift $\Delta E_{2P-2S}$ in muonic hydrogen
depends on the proton charge radius, $R_E$ through
\cite{Pachucki:1996zza}
\begin{eqnarray} 
& & \Delta E_{2P-2S}=205.93(1)\,{\rm meV}- \frac{2}{3}\alpha\pi\phi_{2S}^2(0) R_E^2 + O(\alpha^5), \nonumber \\
\label{eq:Lamb_RE}
\end{eqnarray} 
where the wave function at origin is  given by $\phi_{2S}^2(0)=(\alpha
m_r)^3/8\pi$, $\alpha=e^2/4\pi$ is the fine structure constant, and  $m_r\equiv m_\mu M/(m_\mu+M)$ is the reduced mass with 
$m_\mu$, $M$ the muon and proton masses, respectively. 
The value of the Lamb shift predicted using $R_E$ quoted by 
 the Committee on Data for Science and Technology (CODATA) \cite{CODATA}
\beqn
R_E=0.8768\,(69) {\rm fm},
\label{eq:RE_e}
\eeqn 
 that is  based primarily on the electronic Hydrogen Lamb shift measurement, 
or on the value extracted from the most recent electron scattering data ~\cite{bernauer},
 \beqn
R_E=0.879\,(8)\,\mathrm{fm},
\label{eq:RE_scatt}
\eeqn
differs by $5\sigma$ from the measurement of the muonic hydrogen Lamb
shift by Pohl et al. \cite{pohlmuH}. The later requires a significantly
smaller charge radius,
\beqn
R_E&=&0.84184\,(67)\,\mathrm{fm}.
\label{eq:RE_mu}
\eeqn
  In terms of the Lamb shift, the discrepancy amounts to some 300$\mu$eV that by far exceeds the experimental sensitivity of the muonic experiment \cite{pohlmuH}. The first term in Eq.~(\ref{eq:Lamb_RE}), that represents, up to  $O(\alpha^5)$,  all QED 
 effects associated with the leptonic current is almost three  orders of
magnitude larger than the observed discrepancy. This may lead to a
conclusion that a slight adjustment in one of those terms could 
resolve the whole puzzle. These higher-order QED 
corrections, however,  have been known for  a long time and are well established. The reader is referred to three recent reviews  
 which  assess the full body of  the relevant QED corrections, 
\cite{Pachucki:1996zza,Eides2001,Borie2011}). A non-perturbative numeric evaluation is also
available~\cite{Carroll:2011rv} and yields a similar result, and so does the analysis based in effective non-relelativistic expansion of QED  \cite{Hill:2011wy,Pineda:2011xp}. 
An exotic possibility is a substantial non-universality of lepton-proton interaction, which has  not been
observed before but a more plausible explanation is that  higher order terms in the expansion in $\alpha$ 
 is responsible for the discrepancy. Since QED corrections have a solid founding, attention has been focused
on higher-order, nucleon structure-dependent effects. 
To lowest order, $O(\alpha^5)$ these arise through a two-photon exchange process and potentially 
 bear significant  uncertainty because they involve the complete nucleon excitation spectrum. 

In Section \ref{sec1}, we  assess this two-photon exchange contribution to the Lamb shift using 
forward dispersion relations. Section \ref{sec2} deals with the novel feature
of our approach, were we use the finite energy sum rule (FESR) to relate the value of
the subtraction function that arises in the dispersive calculation to the contribution from the fixed $J=0$ Regge pole. 
Section \ref{sec3} is dedicated to the numerical analysis. Discussion of the results and 
comparison with the existing calculations is summarized in Section \ref{sec4}. 



\section{Dispersion Relations for Compton Scattering} 
\label{sec1}
The $O(\alpha^5)$ contribution to Lamb shift sensitive to proton structure enters through the matrix element of the two-photon exchange (TPE) between the lepton and nucleon integrated over the atomic wave function.
This can be seen as the virtual excitation and de-excitation of the proton by the successive photons, and thus all the complexity of the excited nucleon states is affecting a precision atomic physics computation.
 Taking the standard approach for computing bound state corrections in atomic physics 
 which express nucleon current effects in terms of the atomic wave function at the origin 
  the TPE   contribution to the Lamb shift   is then given by 
   \cite{carlsonvdh,Faustov:1999ga}
\beqn \label{maineq}
E=4\pi i
\frac{\phi_n^2(0)}{2m_l} e^2\!\!\int\frac{d^4q}{(2\pi)^4}
\frac{(q^2+2\nu^2)T_1-(q^2-\nu^2)T_2}{q^4[(q^2/2m_l)^2
-\nu^2]},\nn\\
\eeqn
 where $m_i$, $i=e,\mu$ is the lepton mass in  conventional and muonic
 hydrogen, respectively. The scalar functions $T_{1,2}= T_{1,2}(\nu,q^2)$ with $\nu=(pq)/M$, are the standard   amplitudes that parametrize the  spin-independent hadronic tensor for doubly virtual forward Compton scattering $\gamma^*(q)+N(p)\to\gamma^*(q)+N(p)$, and are given by 
\beqn
T^{\mu\nu}&=&\frac{i}{8\pi M}\int d^4xe^{iqx}\langle N|T[J^\mu(x),J^\nu(0)]|N\rangle\nn\\
&=&
\left(-g^{\mu\nu}+\frac{q^\mu q^\nu}{q^2}\right)T_1(\nu,q^2)\\
&+&\frac{1}{M^2}\left(p^\mu-\frac{pq}{q^2}q^\mu\right)
\left(p^\nu-\frac{pq}{q^2}q^\nu\right)T_2(\nu,q^2), \nn
\eeqn
\indent
The hadronic tensor can be measured in a restricted kinematic range of the variables $\nu$  and $Q^2$ and needs to be extrapolated outside the physical range to compute the integral in Eq.~(\ref{maineq}). The extrapolation is based on analytical continuation.  Specifically, 
 the functions $T_{1,2}$  are discontinuous along the real axis in the complex energy plane $\nu$ with the discontinuity, which is  equal to the imaginary part,  related to the inclusive cross section 
\beqn
{\rm Im}T_{1}(\nu,q^2)&=&\frac{e^2}{4M} F_1\nn\\
{\rm Im}T_{2}(\nu,q^2)&=&\frac{e^2}{4\nu} F_2 \label{TF}\ , 
\eeqn
As customary in dispersive approaches, 
we make use of the complex $\nu=(s-u)/(4M)$ plane. 
Since this variable is crossing-symmetric, upon applying 
Cauchy's theorem, the left and right cut can be combined 
in the same integral, yielding  a relatively simple forward dispersion relation~\cite{Bernabeu:1973zn}, 
\beqn
{\rm Re}\,T_1(\nu,Q^2)&=&T_1(0,Q^2)
+\frac{\nu^2e^2}{2\pi M}{\cal P}\int\limits_{{\nu_{tr}}
}^\infty
d\nu'\frac{F_1(\nu',Q^2)}{\nu'(\nu'^2-\nu^2)}\nn\\
{\rm Re}\,T_2(\nu,Q^2)&=&\frac{e^2}{2\pi}{\cal P}
\int\limits_{\nu_{tr}}^\infty d\nu'\frac{F_2(\nu',Q^2)}{(\nu'^2-\nu^2)},
\eeqn
\indent
While this suffices to reconstruct $T_2$ from knowledge of the dispersive part, $T_1$ 
requires an additional input in the form of a subtraction constant at each $Q^2$, {\it i.e} the function $T_1(0,Q^2)$. This is due to divergence of the unsubtracted dispersive integral at large energies as dictated by the high energy asymptotic properties of the $F_1$ structure function. 
 At the real photon point $Q^2=0$,  the subtraction term is fixed by 
the well-known Thomson-scattering limit, $T_1(0,0) = -\alpha/M$.
For virtual photons however, existing estimates carry large uncertainties. 
They are based on the not so well determined polarizability and the $Q^2$ dependence of elastic form factors.

The  $F_i$ structure functions measured with virtual photons 
receive a contribution from the single nucleon pole (Born terms) at
$\nu_{tr} = \nu_N=\pm Q^2/2M$, and from the unitarity cut 
due to opening of particle production thresholds which start with pion production at 
$\nu_{tr} = \nu_\pi(Q^2) =\pm[ (M+m_\pi)^2-M^2+Q^2]/2M$ (with $m_\pi$ being the pion mass). 
Following~\cite{carlsonvdh}, we divide the contribution to the Lamb
shift into three physically distinct terms that originate 
from the 
subtraction term $T_1(0,Q^2)$, the nucleon pole and finally all
excited intermediate states that may couple to $\gamma N$, respectively 
\beqn
\Delta E&=& \Delta E^{subt} + \Delta E^{el}+ \Delta E^{inel}.  \label{split} 
\eeqn
with 
\beqn
&& \Delta E^{subt}=\frac{\alpha}{m_l}\phi^2_n(0)
\int_0^\infty\frac{dQ^2}{Q^2}\frac{\gamma_1(\tau_l)}{\sqrt\tau_l}T_1(0,Q^2)\nonumber \\
&& \Delta E^{el}=-\frac{\alpha^2 m_l}{M(M^2-m_l^2)}\phi^2_n(0)
\int_0^\infty\frac{dQ^2}{Q^2}\label{lamb} \\
&&\times\left[\left(\frac{\gamma_2(\tau_p)}{\sqrt\tau_p}
-\frac{\gamma_2(\tau_l)}{\sqrt\tau_l}\right)
\frac{G_E^2+\tau_pG_M^2}{\tau_p(1+\tau_p)}\right.\nn\\
&&-\left.\left(\frac{\gamma_1(\tau_p)}{\sqrt\tau_p}
-\frac{\gamma_1(\tau_l)}{\sqrt\tau_l}\right)G_M^2\right] \nn\\
&& \Delta E^{inel}=-\frac{2\alpha^2}{m_lM}\phi^2_n(0)\int_0^\infty\frac{dQ^2}{Q^2}
\int_{\nu_\pi}^\infty\frac{d\nu}{\nu}\nn\\
&&\times\left[\tilde\gamma_1(\tau,\tau_l)F_1(\nu,Q^2)
+\frac{M\nu}{Q^2}\tilde\gamma_2(\tau,\tau_l)F_2(\nu,Q^2)\right], \nn
\eeqn
 $\tau_l=Q^2/(4m_l^2)$, $\tau_p=Q^2/(4M^2)$, $\tau=\nu^2/Q^2$, and
the auxiliary functions defined by 
\beqn
\gamma_1(\tau)&\equiv&(1-2\tau)\sqrt{1+\tau}+2\tau^{3/2}\nn\\
\gamma_2(\tau)&\equiv&(1+\tau)^{3/2}-\tau^{3/2}-\frac{3}{2}\sqrt\tau\nn\\
\tilde\gamma_1(\tau,\tau_l)&\equiv&\frac{\sqrt\tau_l\gamma_1(\tau_l)-\sqrt\tau\gamma_1(\tau)}{\tau_l-\tau}\nn\\
\tilde\gamma_2(\tau,\tau_l)&\equiv&\frac{1}{\tau_l-\tau}\left(
\frac{\gamma_2(\tau)}{\sqrt\tau}-\frac{\gamma_2(\tau_l)}{\sqrt\tau_l}\right).
\eeqn
\indent
Note that generally, besides the integral over the muon 
 continuum that is represented in the above equations, a sum over the discrete spectrum must be
 taken. The latter contributes to the Lamb shift at order ${\cal
   O}(\alpha^6)$ and is dropped from our considerations.
Using these formulae, in ~\cite{carlsonvdh}   
 the inelastic contribution,  $\Delta E^{inel}$ was evaluated using the photo-absorption cross section parametrization of 
 ~\cite{bostedchristy} for the resonance region complemented with the high energy parametrization of~\cite{capella}. 
Their elastic (nucleon-pole)  contribution,  $\Delta E^{el}$ was computed using three different phenomenological 
parametrizations of nucleon electromagnetic form factors
\cite{bernauer, Kelly, Arrington}.
Here we also give an independent evaluation of the two contributions.
  For $\Delta E^{inel}$ we use a recent parametrization of inclusive structure
functions~\cite{gorchtein-gZ} that  also uses the 
 parametrization of the resonance region from  ~\cite{bostedchristy} 
 but it uses a modified Regge-inspired  background that is fitted to the total photoabsorption cross section of 
  \cite{breitweg}. The $Q^2$-dependence is introduced as in ~\cite{alwall}.
  For $E^{el}$, we use the parametrization from \cite{Kelly} to finally obtain
\begin{equation} 
\Delta E^{el} =-30.1\;\mu\mathrm{eV}
, \;\;  \Delta E^{inel} = -13.0 \;\mu\mathrm{eV}
\end{equation} 
Within errors these agree with computation in the original analysis of  \cite{carlsonvdh}
\begin{equation} 
\Delta E^{el} =-29.5\pm 1.3\;\mu\mathrm{eV}, \;\;  \Delta E^{inel} = -12.7\pm 0.5 \;\mu\mathrm{eV}.
\end{equation} 

\section{Evaluation of the subtraction term}
\label{sec2}
\subsection{Finite energy sum rules}
While previous analyses concentrate on the low energy constraints for the subtraction term, here we focus on implications 
 of the high energy behavior for constraining the subtractions.  
  This is done by exploiting the finite energy sum rule (FESR) for
the Compton amplitude. 
The subtraction term in the dispersion relation (DR) for $T_1$
 arises because  the high-energy photo absorption cross section does not vanish asymptotically.  
 It can be well described by a Regge-theory inspired parametrization 
\begin{equation}
\sigma_T  \to 
 \sigma_T^R(\nu,0)  
=c_P(0)  \left(\frac{\nu}{\nu_0}\right)^{\alpha_P-1}
+c_R(0)  \left(\frac{\nu}{\nu_0}\right)^{\alpha_R - 1}.
\label{sigmaR}
\end{equation} 
with the effective Pomeron and leading Regge trajectory intercepts given by $\alpha_P = 1.097$ and $\alpha_R = 0.5$, respectively. The contributing to the cross section is determined by 
$c_P(0) = 68.0 \pm 0.2
 \mu \rm b$ and 
$c_R(0) = 99.0 \pm 1.2
\mu \rm b$, 
 with $\nu_0 = 1$ GeV. 

The corresponding contribution to the Compton amplitude $T_1$ of this Regge part is given by 
\begin{eqnarray}
{\rm Im} T_1^R(\nu,0)=(\nu/4\pi)\sigma_T^R(\nu,0) \\ \nonumber
{\rm Re}\,T_1^R(\nu,0) = 
\frac{\nu^2}{2\pi^2}{\cal P}\int_{0}^\infty
d\nu'\frac{\sigma_T^R(\nu')}{\nu'^2-\nu^2}
\end{eqnarray}

Following \cite{damashek}, we write a dispersion relation for the difference, 
$T_1-T_1^R$,
\beqn \label{DRnoRegge}
&&{\rm Re}\,T_1(\nu,0)-{\rm Re}\,T_1^R(\nu,0)= \nn \\
&&= -\frac{\alpha}{M} +\frac{\nu^2}{2\pi^2}{\cal P}\int_{\nu_\pi}^\infty
d\nu'\frac{ \sigma_T(\nu') - \sigma_T^R(\nu') }{\nu'^2-\nu^2}.
  \label{sub}
\eeqn
With the large-$\nu$ tail thus removed, the dispersion integral on the
right hand side of Eq.~(\ref{sub}) is dominated by energies below  
a scale $N= O(\nu_0)$ which is discussed below. Removal of the asymptotic contribution from the dispersive integral introduces a new subtraction, $C_\infty$ defined by, 
\begin{equation} 
C_\infty(0)\equiv \left.[{\rm Re}\,T_1(\nu,0)-{\rm Re}\,T_1^R(\nu,0)]\right|_{\nu\to\infty} .
\end{equation} 
With the help of currently available high energy data, $C_\infty(0)$  has recently been determined with high accuracy ~\cite{Gorchtein:2011xx} and 
 it follows from Eq.~(\ref{DRnoRegge}) that it is related to the high energy parameters by 
\begin{eqnarray} \label{realFESR}
 C_\infty(0)  & =  &  -\frac{\alpha}{M}  
   -\frac{1}{2\pi^2}\int_{\nu_\pi}^{N}
d\nu'\sigma_T(\nu',0)  \nonumber \\
&+ & \frac{\nu_0}{2\pi^2}\sum_{i=P,R} \frac{c_i(0)}{\alpha_i} \left( \frac{N}{\nu_0}\right)^{\alpha_i} \nonumber \\
\end{eqnarray}
The resonance contribution given by the integral over the photoabsorption cross section 
 is well established  and can be readily evaluated from the low energy data. The
parameter $N$ defines the lowest photon energy above which  Regge
parametrization suffices to describe the data, which in the analysis of ~\cite{Gorchtein:2011xx} was taken to be 2 GeV. 
 From this  analysis it follows that $C_\infty(0)=(-0.72\pm 0.35)\, \mu$b GeV. 

\begin{figure}
\includegraphics[width=6cm]{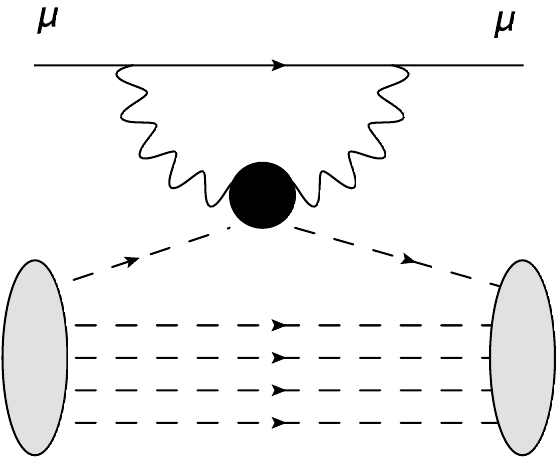}
\caption{{
The residual term for the high energy Compton amplitude corresponding to a fixed pole at $J=0$ complex angular momenttum plane. It corresponds to  Compton scattering on a pointlike quark at instant light-cone time. \label{fixedpole}}
} 
\end{figure}

For our application to muonic hydrogen we need to generalize the above, real Compton amplitude dispersion relation 
 to the virtual photon case. Using the relation 
\beqn
F_1(\nu,Q^2)&=&\frac{M\nu(1-x)}{\pi 
{e^2}\label{F1-sigmat}
}\sigma_T(\nu,Q^2), \label{H} 
\eeqn
where $x=Q^2/(2M\nu)$, we may write
\beqn
T_1(\nu,Q^2)=T_1(0,Q^2)
+\frac{\nu^2 {e^2} 
}{2\pi M}\int\limits_{\nu_\pi(Q^2)}^\infty
\frac{d\nu' F_1(\nu',Q^2)}{\nu'(\nu'^2-\nu^2)}
\eeqn
In analogy  to the real photon case we introduce 
 the Regge-theory motivated representation for the high-energy data valid for 
$\nu\geq N(Q^2)$,
\beqn
{\rm Re}\,T_1^R(\nu,Q^2)=
 \frac{\nu^2
{e^2}
}{2\pi M}{\cal P}\int_{0}^\infty\!\!\!
d\nu'\frac{F_1^R(\nu',Q^2)}{\nu'(\nu'^2-\nu^2)},
\eeqn
with 
\begin{equation} 
F_1^R(\nu,Q^2)=\frac{M\nu_0}{\pi 
{e^2}
}\sum_{i=P,R}c_i(Q^2)\left(\frac{\nu}{\nu_0}\right)^{\alpha_i}. \label{FR} 
\end{equation}
\indent
The generalization of Eq. (\ref{FR}) is not unique since in principle
$\nu_0$ and $\alpha_i$ might be made $Q^2$-dependent. These eventual
$Q^2$-dependences for low $Q^2\lesssim1$ GeV$^2$ that are of interest here 
can however be absorbed in $c_i(Q^2)$ without loss of generality. 
The coefficients $c_i(Q^2)$ must reduce to those found for real 
photons at $Q^2=0$ that are listed below Eq.~(\ref{sigmaR}). 
{Their $Q^2$ dependence, and that of $N(Q^2)$, is obtained by matching the Regge-parametrization of Eq.~(\ref{FR}) and $F_1(\nu,Q^2)$ defined by Eq.~(\ref{H}) 
}
For $\nu\geq N(Q^2)$ and moderate $Q^2\leq1$ GeV$^2$, we obtain 
\beqn
c_P(Q^2)&=&c_P(0) \nn\\
c_R(Q^2)&=&c_R(0)-(20\pm10) \mu \rm b  \left( \frac{Q}{\mbox{GeV}}\right)^2
\end{eqnarray}
and 
\begin{equation} 
N(Q^2)\approx 5\,GeV+\frac{Q^2}{2M}, \label{Reggevalidity}\ .
\end{equation} 
\indent
Note that the presence of the factor $1-x = 1-Q^2/2M\nu$ in
the relation between $\sigma_T$ and $F_1$, Eq.~(\ref{F1-sigmat})
requires a value of $N(Q^2)$ larger than that found for real photons
$N(0)$. In any case, the resulting FESR will not be sensitive to the
value of $N$, as long as the Regge amplitude correctly represents 
the data for all $\nu>N$. The values $c_P(0),c_R(0)$ are fixed by very
precise fit to real photoabsorption data, and $c_P(Q^2)$ is moreover
fixed to its real photon value (for low $Q^2\lesssim1$ GeV$^2$ only) 
to ensure that asymptotically $\sigma_T-\sigma_T^R$ vanishes, the
assumption that is crucial for the FESR method. This effectively
leaves the $Q^2$-slope of the coefficient $c_R(Q^2)$ taken as a linear
function the only parameter that has an uncertainty, and we assign a
generous 50\% uncertainty thereto. The analog of Eq.~(\ref{realFESR}) at finite $Q^2$, 
\beqn
C_\infty(Q^2)\equiv \left.[ {\rm Re}\,T_1(\nu,
Q^2)-{\rm Re}\,T_1^R(\nu, Q^2)]\right|_{\nu \to \infty}
\eeqn
satisfies now 
\beqn
C_\infty(Q^2)&=&T_1(0,Q^2)
-\frac{
{e^2}
}{2\pi M}\int\limits_{\nu_\pi(Q^2)}^{N(Q^2)}
\frac{d\nu'}{\nu'}F_1(\nu',Q^2)\nn\\
&+&\frac{\nu_0}{2\pi^2}\sum_i\frac{c_i(Q^2)}{\alpha_i}\left(\frac{N(Q^2)}{\nu_0}\right) ^{\alpha_i} 
\label{cq} 
\eeqn
It is expected, that at high $Q^2$ $C_\infty(Q^2)$ is finite and represents a light-cone instantaneous, two-photon interaction on a point-like quark ~\cite{Brodsky:1971zh}, as  depicted in figure~\ref{fixedpole}.
This causes no problem in the first of equations~(\ref{lamb}) for $E^{subt}$ that is convergent upon substitution of a constant contribution to $T_1(0,Q^2)$.  The constant $C_\infty(Q^2)$ is related to the virtual Compton amplitude $T_1(0,Q^2)$ through Eq.~(\ref{cq})  and enters the Lamb shift though $E^{subt}$.

\begin{figure}
\includegraphics[width=6cm]{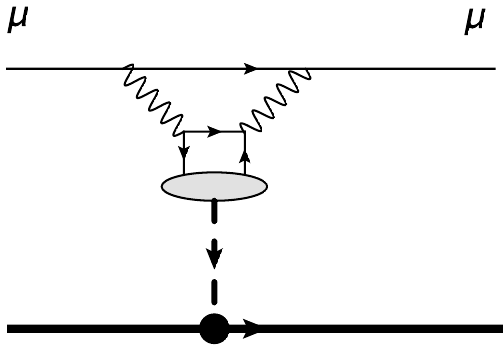}
\caption{{
Regge exchanges in the $t$-channel dominate the high-energy part of the Compton amplitude. 
\label{tchannel}}
} 
\end{figure}

\begin{figure}
\includegraphics[width=6cm]{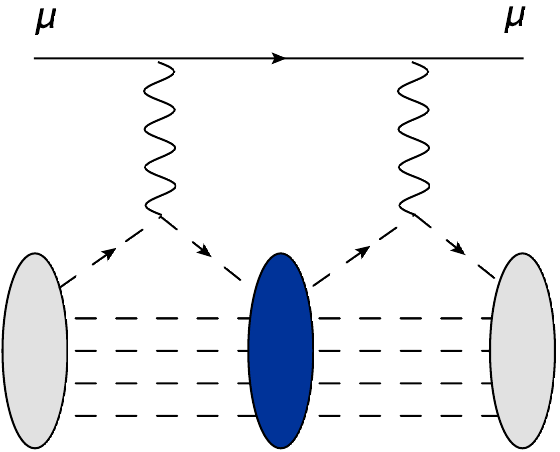}
\caption{{
The low and intermediate energy region is described by a sum over a s-channel resonances that are photoexcitations of the nucleon. 
\label{schannel}}
} 
\end{figure}
To evaluate the integral on the right hand side of Eq.~(\ref{cq}) we need a
parametrization of the virtual photon-proton cross section to substitute in Eq.~(\ref{H}), for which we use the form obtained in 
  \cite{gorchtein-gZ} fits to electroproduction data, 
\beqn
\sigma_T(W^2,Q^2)&=&\sum_a BW_a(W^2) F^2_a(Q^2)\\
&+&
\left[1-e^{\frac{(M+m_\pi)^2-W^2}{M^2}}\right]\sigma_{tot}^R(W^2,0)
F_B(Q^2).\nn
\eeqn
In the  first term the summation runs over nucleon resonances with $BW$ standing for a Breit-Wigner propagator, $BW_a(W^2)$, and electromagnetic transition form factors given by $F_a(Q^2)$. The 
 second term represents a smooth background. 
Expressing $T_1(0,Q^2)$ in terms of the $J=0$ pole contribution, $C_\infty(Q^2)$ yields, 
\beqn \label{atlastwegetthere} 
&&T_1(0,Q^2) =C_\infty(Q^2) 
-\frac{\nu_0}{2\pi^2
}\sum_i\frac{c_i(Q^2)}{\alpha_i}\left(\frac{N(Q^2)}{\nu_0}\right)^{\alpha_i}\nn\\
&&+\frac{1}{2\pi^2
}\int\limits_{\nu_\pi(Q^2)}^{N(Q^2)}
d\nu' \left(1-\frac{Q^2}{2M\nu}\right)\sigma_T(\nu',Q^2)\ ,
\eeqn
which is the main result of this paper. It expresses the low-energy function $T_1(0,Q^2)$ that enters the Lamb shift through  
 $E^{subt}$ in  Eq.~(\ref{split})  in terms of  three distinct contributions  with clear physical interpretation, which are diagrammatically shown in figures~\ref{fixedpole}, \ref{tchannel} and \ref{schannel}.
The last two are the $t$-channel Regge exchanges and $s$-channel resonance contributions; the split between the two 
 is determined by $N(Q^2)$.   The first term is the $J=0$, fixed-pole contribution to virtual Compton scattering $C_\infty(Q^2)$ \cite{Brodsky:2008qu} to which we now turn our attention to. 

\subsection{
{Analysis of the fixed pole} 
}
The $J=0$ fixed pole in Compton scattering was introduced in~\cite{Creutz:1968ds} and studied in phenomenological models 
{\it e.g.} in  \cite{Brodsky:1971zh,Brodsky:2008qu,Brodsky:2007fr,Brodsky:2009bp}. 
 Such an $s$ and $t$ independent contribution has been analyzed in the
 kinematic region where both $-t$, $s$ are large, $s,\,-t\gg M_N^2$ and the existing data in this region~\cite{Shupe:1979vg,Danagoulian:2007gs} supports existence of the fixed pole. 

For real Compton scattering  $C_\infty(0)$ was determined in \cite{Gorchtein:2011xx}, 
 however, in Eq.(\ref{atlastwegetthere}) $C_\infty$ is evaluated at
finite $Q^2$.  Theory suggests that at  asymptotic $Q^2$, $C_\infty(Q^2)$ is constant~\cite{Brodsky:1971zh}, but this has not been experimentally established; it might be so in the future with the help of the Deeply Virtual Compton Scattering program at Jefferson lab. To allow for the possibility of a $Q^2$ dependence, we subtract Eq.~(\ref{realFESR}) (real FESR) from Eq.~(\ref{atlastwegetthere}) (virtual FESR), and changing the integration variable from $\nu$ to $\omega = \nu-Q^2/2M$,  obtain
 \beqn
&&T_1(0,Q^2)=-\frac{\alpha}{M} + [C_\infty(Q^2)-C_\infty(0)]\\
&&+\frac{1}{2\pi^2}\int\limits_{\nu_\pi(0)}^{N(0)}
d\omega\left[\frac{\omega}{\omega+\frac{Q^2}{2M}}\sigma_T(\omega,Q^2)-\sigma_{T}(\omega,0)\right]\nn\\
&&+\frac{\nu_0}{2\pi^2}\! \! \sum_{i=P,R}\! \!
\left[\!
\frac{c_i(0)}{\alpha_i}\!\left(\frac{N(0)}{\nu_0}\right)^{\alpha_i}\!\!\!\!\!\!
-\frac{c_i(Q^2)}{\alpha_i}\!\left(\frac{N(Q^2)}{\nu_0}\right)^{\alpha_i}
\!\!\!\!\!\! F_B(Q^2)\!\right] . \nonumber \\
\label{final} 
\eeqn
This is a rigorous representation of the  subtraction term in the virtual  Compton amplitude. 
If the fixed pole were $Q^2$ independent, as suggested by \cite{Brodsky:1971zh}, $C_\infty$ would 
  drop out of this equation. Since this is not established
  experimentally,  we also provide an order of magnitude estimate under the assumption that $C_\infty(Q^2)$ falls with $Q^2$. 
 
For the estimates of the uncertainty associated with 
$C_\infty(Q^2)-C_\infty(0)$ we use a parametrization 
\beqn
C_\infty(Q^2)-C_\infty(0)=\frac{Q^2}{\Lambda^2+Q^2}[C_\infty(\infty)-C_\infty(0)]\ ,
\eeqn
with a typical scale $\Lambda=1$ GeV and $C_\infty(\infty)=0$.  

\section{Numerical Analysis} 
\label{sec3}
If we substitute Eq.~(\ref{final}) in the expression for $E^{subt}$ in
Eq.~(\ref{lamb}) we see that the result is IR divergent. This is due
to the Thomson term, $T_1(0,0)=-\frac{\alpha}{M}$. Physically, it
corresponds to exchange of soft Coulomb photons that is already taken
into account at the level of atomic wave functions, and has to be
subtracted in order to avoid double-counting. We are left with the
following convergent integral to be evaluated:
\beqn\label{neededintegrals}
\Delta E^{subt}=4\alpha\phi^2_n(0)
\int\limits_0^\infty dQ\gamma_1(\tau_l)
\frac{T_1(0,Q^2)+\frac{\alpha}{M}}{Q^2}\,.
\eeqn
The contribution from $T_1(0,Q^2)$ to the Lamb shift can be written as a sum of several terms,
\beqn
\Delta E^{subt}=
\sum_i \Delta E_i^{res}+
\Delta E^{Back}+\Delta E^{Regge}
\eeqn
We evaluated the respective integrals in Eq.~(\ref{neededintegrals}) numerically. Below, 
 we quote the individual contributions from each of the well-established resonances, the non-resonant background,  and the Regge part, respectively,
 
\beqn \label{contributions}
\Delta E_{\Delta(1232)}&=&(0.95\pm0.09)\,\mu{\rm eV}\nn\\
\Delta E_{S_{11}(1535)}&=&(-4.02\pm3.14)\,\mu{\rm eV}\nn\\
\Delta E_{D_{13}(1520)}&=&(0.41\pm0.09)\,\mu{\rm eV}\nn\\
\Delta E_{S_{11}(1665)}&=&(-0.23\pm0.16)\,\mu{\rm eV}\nn\\
\Delta E_{F_{15}(1680)}&=&(-0.32\pm0.06)\,\mu{\rm eV}\nn\\
\Delta E_{P_{11}(1440)}&=&(0.10\pm0.02) \,\mu{\rm eV}\nn\\
\Delta E_{F_{37}(1950)}&=&(-0.76\pm0.26)\,\mu{\rm eV}\nn\\
\Delta E^{Back}&=&(-29.34\pm2.93)\,\mu{\rm eV}\nn\\
\Delta E^{Regge}&=&(36.55\pm1.6)\,\mu{\rm eV}\,,
\eeqn
Adding the above contributions to the subtraction term, 
\beqn
\Delta
E^{subt}=(3.3\pm4.6)\,{\mu \rm eV}
\eeqn
It can be noted that there are strong cancellations between various
terms. The size of the correction is almost entirely given by the sum
of three contributions, $\Delta E^{Regge},\Delta E^{Back}$ and $\Delta E_{S_{11}(1535)}$. To discuss the uncertainty it thus suffices to constrain the uncertainty in these three contributions. Regge and background
contributions are large, opposite in size and cancel to about
80\%. The background contribution is obtained from a fit to excellent
experimental data over a wide range of $W^2,Q^2$ (see
Ref.\cite{bostedchristy} for a full list of references) and a relative
uncertainty of 10\% is reasonable. The Regge contribution is related to
the background since they are constructed to coincide at high
energies, and assigning an extra uncertainty here would lead to double
counting. We assign a 50\% uncertainty on the $Q^2$-slope of the Reggeon
strength $c_R(Q^2)$. For the resonances, we assign the uncertainties
listed in the PDG \cite{PDG} for the $R\to N\gamma$ transition helicity
amplitudes. The main uncertainty is due to $S_{11}(1535)$, and we
believe that this estimate of uncertainties is very conservative. The
actual fit describes the data in the second resonance region certainly 
better than $\pm$70\%. We believe that this uncertainty can be further
reduced. 

Finally, we obtain for the hadronic ${\cal O}(\alpha^5)$ contribution
to the $2P-2S$ Lamb shift in muonic deuterium set forth in Eq.~(\ref{maineq})
\beqn
\Delta E&=&(-40\pm 5)\,\mu{\rm eV}.
\eeqn

\section{Discussion}
\label{sec4}
We have split the contribution of the nucleon's Compton tensor to the Lamb shift of the muonic hydrogen atom into three parts, 
$E^{el}$, $E^{inel}$ and $E^{subt}$. The first two, corresponding to elastic scattering off the proton and photoexcitation of resonances are in agreement with previous work by other authors. The last term contains the contribution of the real subtraction to the Compton tensor and is the only one where significant uncertainty has remained.  Specifically, in the analyses of  \cite{Pachucki:1996zza} the subtraction function was identified with 
\beqn
T_1(0,Q^2)&=&-\frac{\alpha}{M}F_D^2(Q^2)+Q^2\beta(Q^2),
\eeqn
where $F_D(Q^2)$ stands for the Dirac form factor, and $\beta(Q^2)$
for the generalized magnetic polarizability that for real photons
reduce to the usual magnetic polarizability of Compton scattering,
$\beta(0)=\beta_M$. Its $Q^2$ dependence was taken by analogy with elastic form factors. In Ref. \cite{carlsonvdh} it was argued that 
\beqn
T_1(0,Q^2)&=&-\frac{\alpha}{M}+Q^2\beta(Q^2),
\eeqn
where we put together the two contributions identified in
\cite{carlsonvdh} as $T_1^{NB}(0,Q^2)=Q^2\beta(Q^2)$ and
$T_1^{B,\,no-pole}(0,Q^2)=-\frac{\alpha}{M}$ for clarity.
The common feature of the two approximations is that at $Q^2=0$ they
reduce to the Thomson term. However, they differ already in the first
derivative, and they effectively operate with two different values of
$\beta$ that is a measured quantity. We define 
\beqn
\bar T_1(Q^2)\equiv\frac{T_1(0,Q^2)+\frac{\alpha}{M}}{Q^2},
\eeqn
the function that enters the calculation of the Lamb shift, 
and evaluate this function at $Q^2=0$. With the model of
 \cite{Pachucki:1996zza} one obtains 
\beqn
\bar T_1(0)=-\frac{\alpha}{M}2F'_D(0)+\beta,
\eeqn
while the model of Ref. \cite{carlsonvdh} gives
\beqn
\bar T_1(0)=\beta.
\eeqn
\indent
The difference is not small and amounts to 3.3$\times10^{-4}$ fm$^3$,
of the same size as the polarizability  itself. 
What complicates the issue is
the impossibility to measure $T_1(0,Q^2)$ directly since the
kinematical arguments are in the unphysical region. The problem of
low-energy expansion of doubly virtual Compton scattering was
approached  by two of us in \cite{LEXsumrules} in terms of a fully model-independent
low-energy theorem. It was found that it is only possible
to unambiguously identify $T_1(0,Q^2)$ with a combination of known or
measurable quantities (form factors and polarizabilities) modulo a 
dispersion integral in the annihilation channel that is largely
unknown. Rewriting the findings of Ref. \cite{LEXsumrules} for
$T_1(0,Q^2)$ we find
\beqn
T_1(0,Q^2)&=&-\frac{\alpha}{M}[F_D^2(Q^2)-\tau F_P^2(Q^2)]
+Q^2\beta(Q^2)+\dots,\nn\\\label{LEX}
\eeqn
where we omitted terms coming from that dispersion
integral in the annihilation channel. 
The reason for such detailed discussion is to remind the reader that
to relate the {\it unphysical} subtraction constant $T_1(0,Q^2)$ to
{\it measurable} quantities like the polarizability and elastic form
factors, a good deal of caution should be exercised. 

\begin{figure}
\includegraphics*[width=3.3in]{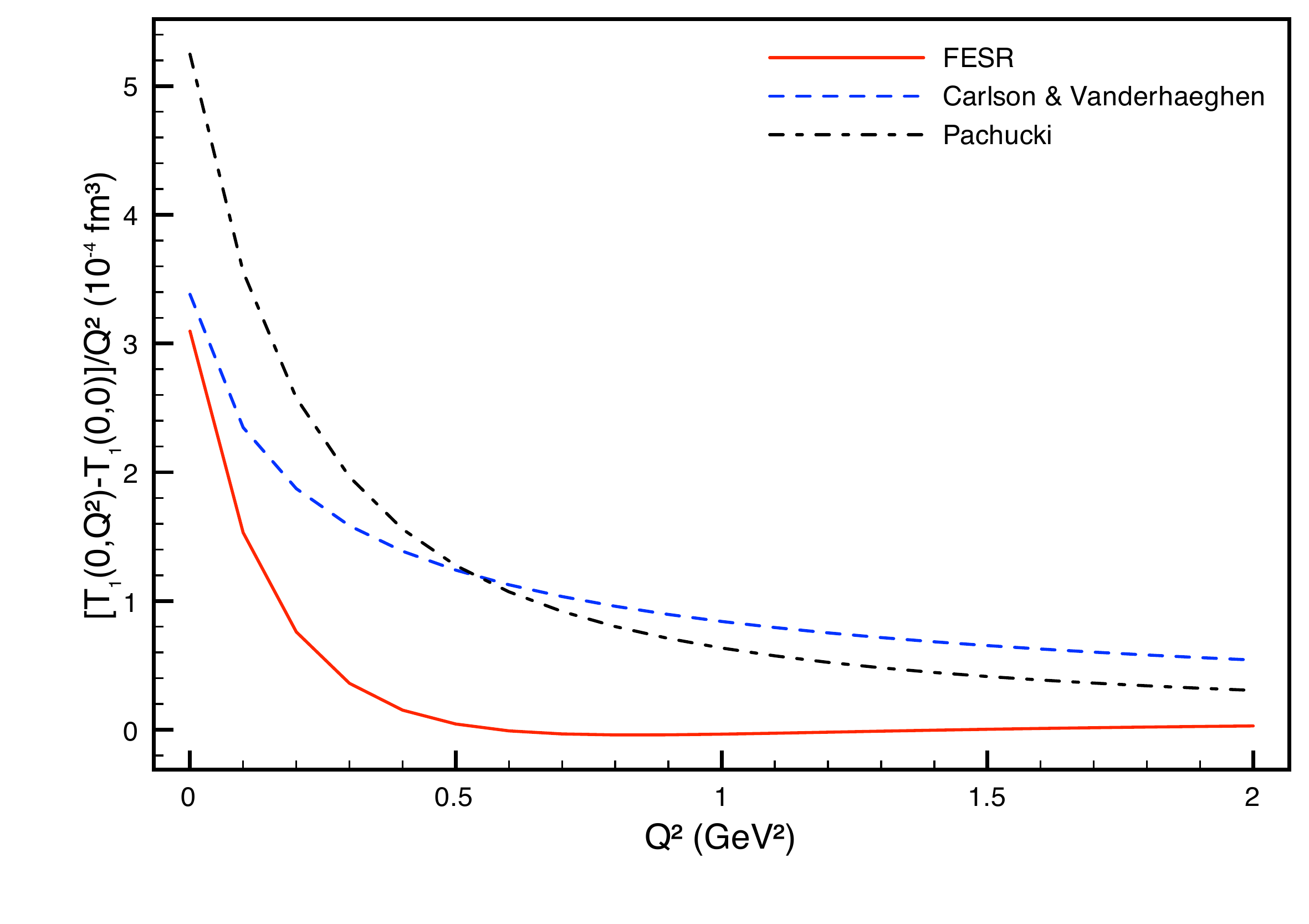}
\caption{
{(Color online) Subtraction function $[T_1(0,Q^2)-T_1(0,0)]/Q^2$ in
  units of $10^{-4}$ fm$^3$ as obtained from FESR (solid), from the
  model of Ref. \cite{carlsonvdh} (dashed) and from
  Ref. \cite{Pachucki:1996zza} (dash-dotted). 
\label{fig:SubConst}}
}
\end{figure}

Following the analysis presented in this paper, the systematic uncertainty in the Lamb shift from this term has been significantly reduced. 
We have employed the method of the Finite Energy Sum Rules to analyze this term, explicitly displaying the contributions it receives from the known $t$-channel Regge and $s$-channel resonances. There is no double counting of these resonances with respect to $E^{inel}$. The alternative analysis presented here provides information on the subtraction
term from Regge theory and the resonance region, reducing the unknowns to the fixed pole of Compton scattering. Our Finite
Energy Sum Rule in Eq.~(\ref{final}) has for the first time made it 
possible to predict the $Q^2$-dependence of the subtraction function directly from 
existing experimental data. In Fig. \ref{fig:SubConst} we compare the function $\bar T_1(Q^2)$ as
obtained from FESR to phenomenological {\it Ans\"atze} of previous
analyses. We observe that all approaches effectively have similar values
of $\bar T_1(0)$ but in view of the complicated situation with the
low-energy theorem discussed above we stress that this is a
coincidence. Neglecting the $t$-channel contributions in
Eq. (\ref{LEX}) and removing the contributions of the form factors
(3.3$\times10^{-4}$ fm$^3$ and 1.5$\times10^{-4}$ fm$^3$) we would
arrive at $\beta=-1.8\times10^{-4}$ fm$^3$. 

We have shown that the contribution of the 
subtraction term $\Delta E^{subt}$ is small, $\approx3\mu$eV, and its large relative error 
of order 5$\mu$eV does not alter the conclusion that the overall 
contribution of the nucleon photoexcitation processes to the Lamb
shift in muonic hydrogen is about
-40$\pm$5$\mu$eV. This is in agreement with the recent evaluation of Carlson and Vanderhaeghen  
($-37\pm 2.5\mu$eV), both being somewhat larger than earlier determinations of order $-20\mu$eV.
Our overall estimated uncertainty has increased a bit respect to earlier work~\cite{Pachucki:1996zza} as well as chiral perturbation theory~\cite{Nevado:2007dd}, but we feel we have better control of systematic unknowns.

The 300$\mu$ eV discrepancy between the direct muonic Hydrogen 
Lamb shift measurement and estimates for it based on
usual (electronic) Hydrogen is unnaturally large for the hadronic
structure-dependent corrections at order ${\cal O}(\alpha^5)$ that have been proposed in the literature, basically Eq.~(\ref{maineq}), and the
explanation must be looked for elsewhere.

\acknowledgments
{
We thank Stan Brodsky for suggesting that we reexamine this issue and various useful comments. 
Work supported by US DOE grant DE-FG0287ER40365, German SFB 1044 and Spanish grants FPA2011-27853-01 and FIS2008-01323.
}



\begin{thebibliography}{90}

\bibitem{Jentschura:2010ej} 
  U.~D.~Jentschura,
  Annals Phys.\  {\bf 326}, 500 (2011)
  [arXiv:1011.5275 [hep-ph]];
  U.~D.~Jentschura,
  Annals Phys.\  {\bf 326}, 516 (2011)
  [arXiv:1011.5453 [hep-ph]].



\bibitem{Pachucki:1996zza} 
  K.~Pachucki,
  Phys.\ Rev.\ A {\bf 53}, 2092 (1996).

\bibitem{CODATA} P. J. Mohr, B. N. Taylor, and D. B. Newell,
  Rev. Mod. Phys. {\bf 80}, 633 (2008).

\bibitem{bernauer} J. C. Bernauer et al. (A1), Phys. Rev. Lett. {\bf 105}, 242001
(2010).

\bibitem{pohlmuH} R.~Pohl et al., Nature News doi:10.1038/news.2010.337 ;
R. Pohl {\it et al.}, Nature {\bf  466} 213 (2010). A recent reevaluation puts the discrepancy at a higher 7$\sigma$, see A. Antognini {\it et al.}, Science {\bf 339}, 417 (2013).


\bibitem{Eides2001} M.~I.~Eides, H.~Grotch, and V.~A.~Shelyuto,
  Phys.~Rept.~{\bf 342}, 63 (2001), hep-ph/0002158.

\bibitem{Borie2011} E.~Borie, Annals Phys. {\bf 327} (2012) 733.

\bibitem{Carroll:2011rv} 
  J.~D.~Carroll, A.~W.~Thomas, J.~Rafelski and G.~A.~Miller,
  Phys.\ Rev.\ A {\bf 84}, 012506 (2011)
  [arXiv:1104.2971 [physics.atom-ph]].

\bibitem{Hill:2011wy} 
  R.~J.~Hill and G.~Paz,
  Phys.\ Rev.\ Lett.\  {\bf 107}, 160402 (2011)
  [arXiv:1103.4617 [hep-ph]].

\bibitem{Pineda:2011xp} 
  A.~Pineda,
  arXiv:1108.1263 [hep-ph].


\bibitem{carlsonvdh} Carl.~E.~Carlson, Marc Vanderhaeghen,
  Phys.~Rev.~A{\bf84} (2011), 020102. 

\bibitem{Faustov:1999ga} 
  R.~N.~Faustov and A.~P.~Martynenko,
  Phys.\ Atom.\ Nucl.\  {\bf 63}, 845 (2000)
  [Yad.\ Fiz.\  {\bf 63}, 915 (2000)]
  [hep-ph/9904362].




{
\bibitem{Bernabeu:1973zn} 
  J.~Bernabeu and C.~Jarlskog,
  Nucl.\ Phys.\ B {\bf 60}, 347 (1973).
}

\bibitem{bostedchristy} P.E. Bosted, M.E. Christy, Phys.Rev. C81
  (2010) 055213.
\bibitem{capella} Capella et al., Phys. Lett. B337, 358 (1994).
\bibitem{Kelly} J.~J.~Kelly, Phys.~Rev.~C{\bf 70}, 068202 (2004).
\bibitem{Arrington} J. Arrington, W. Melnitchouk, and J. A. Tjon, Phys. Rev.
C{\bf 76}, 035205 (2007), 0707.1861.
\bibitem{gorchtein-gZ} M. Gorchtein, C.J. Horowitz,
  M.J. Ramsey-Musolf, Phys.Rev. C84 (2011) 015502.
  \bibitem{breitweg} J. Breitweg et al., Eur. Phys. J. C 7 (1999), 609.
\bibitem{alwall} J. Alwall, G. Ingelman, Phys. Lett. B 596 (2004), 77.

\bibitem{LEXsumrules} M.~Gorchtein, A.~P.~Szczepaniak,
  Phys.~Rev.~Lett. {\bf 101}, 141601 (2008), 0807.3791

\bibitem{damashek} M.~Damashek and F.~ J.~Gilman, Phys.~Rev.~D {\bf 1}, 1319
(1970).
{
\bibitem{Gorchtein:2011xx} 
  M.~Gorchtein, T.~Hobbs, J.~T.~Londergan and A.~P.~Szczepaniak,
Phys.~Rev.~C{\bf 84}, 065202 (2011). 
  arXiv:1110.5982 [nucl-th].
}

\bibitem{Brodsky:2008qu} 
  S.~J.~Brodsky, F.~J.~Llanes-Estrada and A.~P.~Szczepaniak,  Phys.\ Rev.\ D {\bf 79}, 033012 (2009). 

\bibitem{Brodsky:1971zh} 
  S.~J.~Brodsky, F.~E.~Close and J.~F.~Gunion,
  Phys.\ Rev.\ D {\bf 5}, 1384 (1972).
\bibitem{Creutz:1968ds} 
  M.~J.~Creutz, S.~D.~Drell and E.~A.~Paschos,
  Phys.\ Rev.\  {\bf 178}, 2300 (1969).
\bibitem{Brodsky:2007fr} 
  S.~J.~Brodsky, F.~J.~Llanes-Estrada and A.~P.~Szczepaniak,
  eConf C {\bf 070910}, 149 (2007)
  [arXiv:0710.0981 [nucl-th]].
\bibitem{Brodsky:2009bp} 
  S.~J.~Brodsky, F.~J.~Llanes-Estrada, J.~T.~Londergan and A.~P.~Szczepaniak,
  arXiv:0906.5515 [hep-ph].
\bibitem{Shupe:1979vg} 
  M.~A.~Shupe, R.~H.~Milburn, D.~J.~Quinn, J.~P.~Rutherfoord, A.~R.~Stottlemyer, S.~S.~Hertzbach, R.~RKofler and F.~D.~Lomanno {\it et al.},
  Phys.\ Rev.\ D {\bf 19}, 1921 (1979).
\bibitem{Danagoulian:2007gs} 
  A.~Danagoulian {\it et al.}  [Hall A Collaboration],
  Phys.\ Rev.\ Lett.\  {\bf 98}, 152001 (2007)
  [nucl-ex/0701068 [NUCL-EX]].

\bibitem{PDG}  J.~Beringer et al. (Particle Data Group),
  Phys.~Rev.~D{\bf 86}, 010001 (2012).

\bibitem{Nevado:2007dd} 
  D.~Nevado and A.~Pineda,
  Phys.\ Rev.\ C {\bf 77}, 035202 (2008)
  [arXiv:0712.1294 [hep-ph]].

\end{thebibliography}
\end{document}